\documentclass[preprint,12pt]{elsarticle}

\usepackage{amssymb}
\usepackage{amsmath}
\usepackage{graphicx}
\usepackage{booktabs}
\usepackage{multirow}

\journal{Neural Networks}

\begin{document}

\begin{frontmatter}



\title{Collaborative Bi-Aggregation for Directed Graph Embedding}


\author[label1]{Linsong Liu}
\ead{1221045504@njupt.edu.cn}
\author[label1]{Kejia Chen\corref{correspondingauthor}}
\cortext[correspondingauthor]{Corresponding author}
\ead{chenkj@njupt.edu.cn}
\author[label1]{Zheng Liu}
\ead{zliu@njupt.edu.cn}

\affiliation[label1]{organization={School of Computer Science},
          addressline={Nanjing university of posts and telecommunications}, 
          city={Nanjing},
          postcode={210023}, 
          state={Jiangsu},
          country={China}}

\begin{abstract}
Directed graphs model asymmetric relationships between nodes and research on directed graph embedding is of great significance in downstream graph analysis and inference. Learning source and target embedding of nodes separately to preserve edge asymmetry has become the dominant approach, but also poses challenge for learning representations of low or even zero in/out degree nodes that are ubiquitous in sparse graphs. In this paper, a collaborative bi-directional aggregation method (COBA) for directed graphs embedding is proposed by introducing spatial-based graph convolution. Firstly, the source and target embeddings of the central node are learned by aggregating from the counterparts of the source and target neighbors, respectively; Secondly, the source/target embeddings of the zero in/out degree central nodes are enhanced by aggregating the counterparts of opposite-directional neighbors (i.e. target/source neighbors); Finally, source and target embeddings of the same node are correlated to achieve collaborative aggregation. Extensive experiments on real-world datasets demonstrate that the COBA comprehensively outperforms state-of-the-art methods on multiple tasks and meanwhile validates the effectiveness of proposed aggregation strategies.

\end{abstract}


\end{frontmatter}


\section{Introduction}
\label{}
Graph embedding usually refers to learning a low-dimensional vector containing both the attribute features and the structural features of each node in the graph. The embedding result can facilitate downstream tasks such as link prediction, node classification, graph reconstruction, etc. 

Graphs modeling real-world network systems are often directed, that is, the relations between nodes are asymmetric. However, implicit information from edge directions is often ignored in most graph embedding methods~\cite{grover2016node2vec,hamilton2017inductive,perozzi2014deepwalk,tang2015line}. Although these methods can be extended to handle directed graphs, it is hard to predict edge directions in tasks such as link prediction and graph reconstruction by learning a single representation of nodes. Ideally, directed graph embedding (DGE) can preserve not only the proximity between nodes but also the asymmetry of the proximity.

To tackle this challenge, recent DGE works~\cite{khosla2019node,kollias2022directed,ou2016asymmetric,sun2019atp,zhou2017scalable,zhu2021adversarial} use two embeddings to represent a node. Source embedding represents the node as a source node, containing the structural information of outgoing edges. Target embedding represents the node as a target node, containing the structural information of incoming edges.

Real-world directed networks are often sparse where many nodes do not have sufficient source neighbors or target neighbors for training, resulting in the performance degradation of most DGE methods~\cite{kollias2022directed,ou2016asymmetric,sun2019atp,zhou2017scalable}. This problem was somehow alleviated with two recently developed methods, NERD~\cite{khosla2019node} and DGGAN~\cite{zhu2021adversarial}. The former increases the probability of low-degree nodes being sampled and and the latter generates fake source and target neighbor nodes from a shared latent distribution. However, these two methods as well as other DGE methods do not take the correlation between two types of embeddings into consideration. Although source embedding and target embedding describe two structural features of a node, they are supposed to be intrinsically connected since they describe the same node.

In this paper, a directed graph embedding method based on collaborative bi-directional aggregation, called COBA, is proposed. Firstly, the method extends spatial-based graph convolution from undirected graphs to directed graphs through a bi-directional aggregation strategy that updates source/target embeddings by aggregating the counterparts from source/target neighbors, respectively. Secondly, a reverse aggregation strategy is proposed to handle nodes with zero in/out degrees, which aggregates the counterparts from neighbors in the opposite direction when updating either the source or target embeddings of the central node. Finally, a collaborative aggregation strategy is further proposed to leverage the correlation between source embedding and target embedding of the same node, that is, the target/source embedding of the central node is additionally used when updating its source/target embedding. The proposed method can achieve promising representations for directed graphs through the above aggregation strategy.

The main contributions in this paper can be summarized as follows:
\begin{itemize}
	\item Spatial-based graph convolution is introduced into directed graph embedding for the first time to learn dual embeddings (i.e., source and target embeddings) of nodes.
	\item A collaborative bi-directional aggregation strategy is proposed not only to facilitate learning zero in/out-degree nodes, but also to leverage the correlation between two embeddings.
	\item Comparative experiments on real-world datasets show that COBA comprehensively outperforms state-of-the-art DGE methods on different downstream tasks.
\end{itemize}

\section{Related Work}
Traditional graph embedding research is oriented towards undirected graphs and mainly uses three types of graph modeling methods: matrix factorization~\cite{cao2015grarep,wang2017community}, random walks ~\cite{grover2016node2vec,perozzi2014deepwalk,tang2015line} and deep graph neural networks~\cite{hamilton2017inductive,kipf2016semi,velickovic2017graph,wang2016structural}. Among them, deep graph neural networks represented by spectral-based graph convolutional networks~\cite{kipf2016semi} and spatial-based graph convolutional networks~\cite{hamilton2017inductive,velickovic2017graph} have made remarkable progress and become the prevailing backbone of deep graph learning models.

In the community of directed graph embedding, existing methods can be roughly classified into two categories depending on whether it represents a node with one or two embeddings. Single-embedding methods are mainly based on GNN models. DGCN~\cite{ma2019spectral}, DiGCN~\cite{tong2020digraph} and DGCN~\cite{tong2020directed} extend spectral-based graph convolution to directed graphs but fail to preserve asymmetry directions of edges. FastMap-D~\cite{gopalakrishnan2020embedding}, GREED~\cite{madhavan2020directed} and Gravity-VAE~\cite{salha2019gravity} can preserve asymmetry proximity by defining a sophisticated asymmetric operation instead of inner product but they are time-consuming due to pairwise distance computations.

Recently, most DGE methods tend to learn two embeddings for each node to better represent edge asymmetries. HOPE~\cite{ou2016asymmetric} and ATP~\cite{sun2019atp} factorize a matrix as the source and target embedding matrices of nodes. The matrix in HOPE preserves asymmetric transitivity by approximating high-order proximity and the matrix in ATP incorporates graph hierarchy and reachability information. APP~\cite{zhou2017scalable} and NERD~\cite{khosla2019node} are random walk-based methods that use learned vertex embeddings as source embeddings and context embeddings as target embeddings. DGGAN~\cite{zhu2021adversarial} and DiGAE~\cite{kollias2022directed} are GNN-based methods. The former introduces GAN into directed graph and learns source and target embeddings separately with the use of the fake neighbors generated by two connected generators and the latter employs parameterized GCN layers as encoder and the inner product of the source and target embeddings as decoder to learn two latent representations.

In the above DGE methods, only NERD~\cite{khosla2019node} and DGGAN~\cite{zhu2021adversarial} notice and try to solve the low-degree node embedding problem. NERD~\cite{khosla2019node} increases the sampling probability of low-degree nodes through an alternating random walk strategy. But it fails to update the source embedding for low-out-degree nodes and the target embedding for low-in-degree nodes as these two types of nodes are sampled as target nodes and source nodes, respectively. DGGAN~\cite{zhu2021adversarial} connects two generators by a shared latent variable to jointly generate negative samples but ignores the correlation between source and target embeddings.

This paper proposes a simple yet effective dual-embedding DGE method, which designs a collaborative bi-directional aggregation strategy based on spatial-based graph convolution to solve the source/target embedding problem of low (mostly zero) out/in-degree nodes. To our best knowledge, only Bi-GCN~\cite{bian2020rumor} adopts a bi-directional processing approach that is somewhat related to our work. However, Bi-GCN deals with a tree structure (i.e., rumor tree) not a graph structure, using two spectral GCNs to simulate rumor propagation and dispersion in two directions. Moreover, COBA can adjust the aggregation strategy according to the local structure of nodes, which is essentially different from Bi-GCN.

\section{Preliminaries}

\subsection{Directed Graph}
Given a directed graph $\mathcal{G} = \left \{ \mathcal{V}, \mathcal{E} \right \}$, where $\mathcal{V}$ is the set of nodes and $\mathcal{E}$ is the set of directed edges. Let $N = \left| \mathcal{V} \right|$ be the number of nodes and $M = \left| \mathcal{E} \right|$ the number of directed edges. Each node pair $\left( u, v \right) \in \mathcal{E} \left( u, v \in \mathcal{V} \right)$ denotes a directed edge from $u$ to $v$. $\mathbf{A} \in \mathbb{R}^{N \times N}$ is the adjacency matrix of $\mathcal{G}$. If $\left( u, v \right) \in \mathcal{E}$, then $\mathbf{A} \left( u, v \right) = 1$; otherwise $\mathbf{A} \left( u, v \right) = 0$. $\mathbf{X} \in \mathbb{R}^{N \times F}$is the feature matrix of all nodes in $\mathcal{G}$, where $F$ is the dimension of the feature vector $x_{v}$. Let $\mathcal{N} \left( v \right)^{+} = \left\{ u|\left( u, v \right) \in \mathcal{E} \right\}$ be the set of source neighbors pointing to node $v$ through incoming edges and $\mathcal{N} \left( v \right)^{-} = \left\{ u|\left( v, u \right) \in \mathcal{E} \right\}$ the set of target neighbors pointed by node $v$ through outgoing edges.

\subsection{Dual Embedding for Directed Graphs}
The dual-embedding model in this paper is specifically designed for directed graphs which aims to seek two functions: $f_{s} \left( \mathbf{X}, \mathbf{A} \right) \rightarrow S \in \mathbb{R}^{N \times d}$ and $f_{t} \left( \mathbf{X}, \mathbf{A} \right) \rightarrow T \in \mathbb{R}^{N \times d}$ by optimization, where $d$ is the dimension of node embeddings. Here, each node $v$ in directed graph $\mathcal{G}$ is represented by a source embedding $s_{v} \in \mathbb{R}^{1 \times d}$ and a target embedding $t_{v} \in \mathbb{R}^{1 \times d}$.

\subsection{Spatial-based Graph Convolution}
Spectral-based graph convolution requires the Laplacian matrix to be symmetric, which inherently conflicts with the asymmetry of directed graphs, while spatial-based graph convolution can effectively aggregate information from neighbors through asymmetric directed edges, which is more convenient for DGE tasks.

In GNN models, spatial-based graph convolution represented by GraphSAGE~\cite{hamilton2017inductive} was initially designed for undirected graphs. The representation of central node $v$ is updated by aggregating the information of the sampled neighbors. This paper for the first time introduces spatial-based graph convolution into DGE in a dual embedding framework.

\section{Model}
This section details the collaborative bi-aggregation strategy in our DGE model COBA.

\begin{figure*}[t]
	\centering
	\includegraphics[width=1\columnwidth]{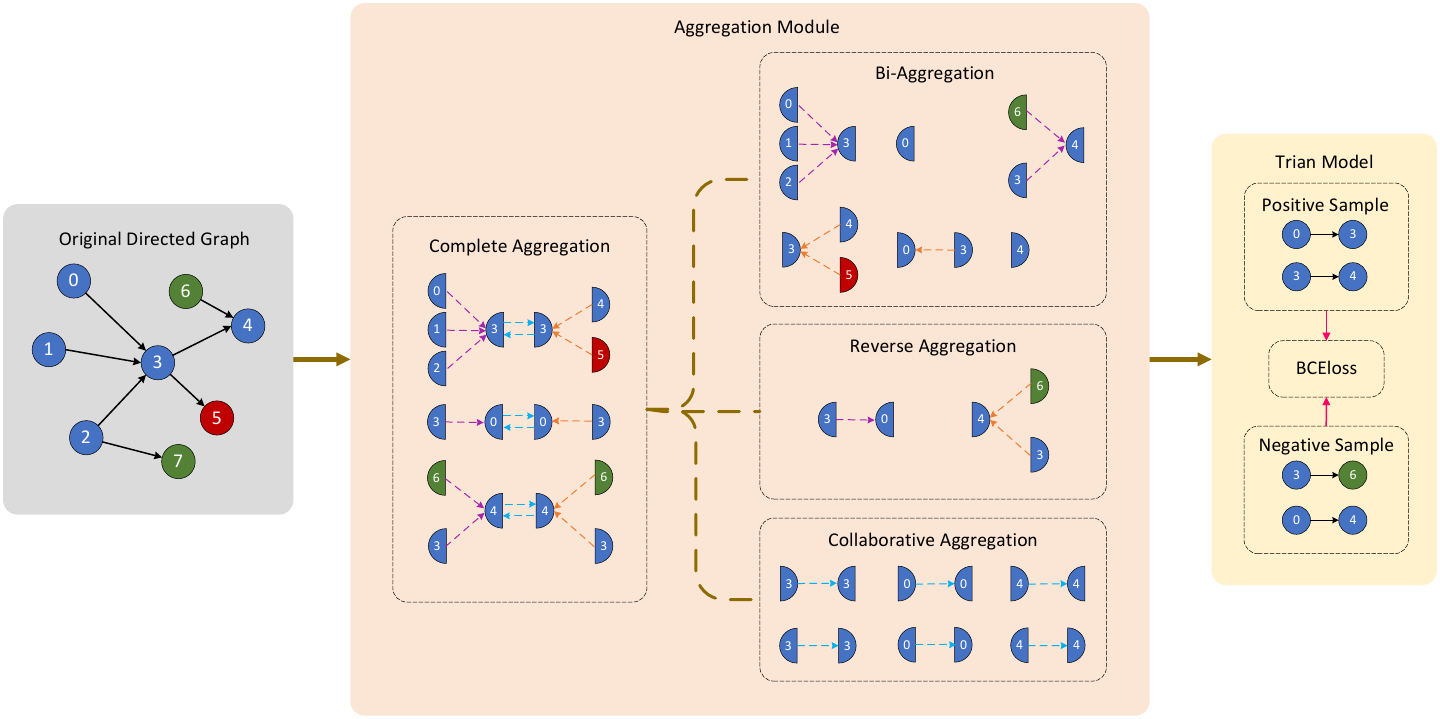}
	\caption{The overall framework of COBA. The colors of nodes represent their different labels. The dotted lines in the aggregation module in the middle indicate different types of aggregation. Left and right semicircles represent source and target embeddings of the node, respectively.}
	\label{figure1}
\end{figure*}

\subsection{Overall Framework}
The overall framework of COBA is illustrated in Figure~\ref{figure1}. The aggregation module in the middle is the crucial part of our model, which contains three types of operations, i.e., bi-aggregation, reverse aggregation and collaborative aggregation. Bi-aggregation is short for bi-directional aggregation, indicating that the central node not only obtains information from source neighbors when updating its source embedding but also from target neighbors when updating its target embedding. Unfortunately, this strategy does not work for zero in-/out-degree nodes since they have no source/target neighbors. Therefore, reverse aggregation is further proposed to address this problem, where zero in-/out-degree nodes can get information from neighbors in the opposite direction (i.e., target/source neighbors). To correlate two embeddings of the same node, collaborative aggregation is finally proposed, that is, the two embeddings of the same node are aggregated with each other.

\subsection{Bi-Aggregation}
Considering the direction consistency of propagation, the property of the central node as a source node is supposed to be similar to that of its source neighbors and analogously, the property of the central node as a target node is supposed to be similar to that of its target neighbors. Therefore, two one-way aggregators, denoted by ${\rm AGG}^{s}$ and ${\rm AGG}^{t}$, are designed to aggregate the source and target embedding of the central node, respectively. For node $v$, its source embedding and target embedding are both initialized by $x_{v}$. When updating the source embedding of node $v$ at layer $l$ (i.e., $s^{\left( l \right)}_{v}$ in Eq.~\ref{eq2}), the source embedding of central node at layer $l-1$ and source embeddings from source neighbors (see Eq.~\ref{eq1}) are all used.
\begin{equation}
	s^{\left( l \right)}_{\mathcal{N} \left( v \right)} = {\rm AGG}^{s} \left( \left\{ s^{\left( l-1 \right)}_{u}, \forall u \in \mathcal{N} \left( v \right)^{+} \right\} \right),
	\label{eq1}	
\end{equation}%
\begin{equation}
	s^{\left( l \right)}_{v} = \sigma \left( {\rm COMBINE} \left( s^{\left( l-1 \right)}_{v}, s^{\left( l \right)}_{\mathcal{N} \left( v \right)} \right) \cdot W^{s} \right).
	\label{eq2}	
\end{equation}%
Likewise, the target embedding of central node $v$ is aggregated as:
\begin{equation}
	t^{\left( l \right)}_{\mathcal{N} \left( v \right)} = {\rm AGG}^{t} \left( \left\{ t^{\left( l-1 \right)}_{u}, \forall u \in \mathcal{N} \left( v \right)^{-} \right\} \right),
	\label{eq3}	
\end{equation}%
\begin{equation}
	t^{\left( l \right)}_{v} = \sigma \left( {\rm COMBINE} \left( t^{\left( l-1 \right)}_{v}, t^{\left( l \right)}_{\mathcal{N} \left( v \right)} \right) \cdot W^{t} \right),
	\label{eq4}	
\end{equation}%
where ${\rm AGG}^{s}$ and ${\rm AGG}^{t}$ are mean aggregators, ${\rm COMBINE}(\cdot,\cdot)$ is the concatenate operator, $W^{s} \in \mathbb{R}^{2d \times d}$ and $W^{t} \in \mathbb{R}^{2d \times d}$ are learnable parameter matrices.

\subsection{Reverse Aggregation}
However, a pure bi-directional aggregation does not perform well for zero in/out-degree nodes. Taking a node with zero in-degree as an example (the situation is similar for nodes with zero out-degree), its source embedding cannot learn neighborhood proximity due to the lack of source neighbors, but can only aggregate itself. A similar problem exists when updating the target embedding of zero out-degree nodes.

This paper proposes a reverse aggregation method to handle this problem. Note that the source and target neighbors of a node are close in the context path (that is, a 2-hop neighborhood), so the node with insufficient source/target neighbors can also obtain additional information from its reverse neighbors when updating itself. To avoid introducing too much noisy data that is not beneficial for identifying edge directions, this paper aggregates the information from reverse neighbors only for zero in/out-degree nodes. Therefore, the bi-aggregation is further improved by optionally using the reverse aggregation according to the degree of nodes.
\begin{equation}
	s^{\left( l \right)}_{\mathcal{N} \left( v \right)} = {\rm AGG}^{s} \left( \left\{ s^{\left( l-1 \right)}_{u}, \forall u \in \mathcal{N} \left( v \right)^{-} \right\} \right),v \in \mathcal{V}^{in}
	\label{eq5}	
\end{equation}%

\begin{equation}
	t^{\left( l \right)}_{\mathcal{N} \left( v \right)} = {\rm AGG}^{t} \left( \left\{ t^{\left( l-1 \right)}_{u}, \forall u \in \mathcal{N} \left( v \right)^{+} \right\} \right),v \in \mathcal{V}^{out}
	\label{eq6}	
\end{equation}%
where $\mathcal{V}^{in}$ and $\mathcal{V}^{out}$ are the set of zero in-degree nodes and zero out-degree nodes, respectively.
\subsection{Collaborative Bi-Aggregation}
The above strategy efficiently learns dual embeddings of nodes including zero in/out-degree nodes. However, it is noticed that two types of embeddings propagate independently on the network without any communication between them. We believe that the source embedding and the target embedding of a common node are latently correlated although they describe different structural properties of the node.

To this end, a collaborative bi-aggregation module is finally assembled. Eq.~\ref{eq2} and ~\ref{eq4} are correspondingly rewritten as: 
\begin{equation}
	s^{\left( l \right)}_{v} = \sigma \left( {\rm COMBINE} \left( s^{\left( l-1 \right)}_{v}, t^{\left( l-1 \right)}_{v}, s^{\left( l \right)}_{\mathcal{N} \left( v \right)} \right) \cdot W^{s} \right),
	\label{eq7}	
\end{equation}%
\begin{equation}
	t^{\left( l \right)}_{v} = \sigma \left( {\rm COMBINE} \left( t^{\left( l-1 \right)}_{v}, s^{\left( l-1 \right)}_{v}, t^{\left( l \right)}_{\mathcal{N} \left( v \right)} \right) \cdot W^{t} \right).
	\label{eq8}	
\end{equation}%

\subsection{Train}
In the model training phase, the positive sample set $\mathcal{E}^{pos}$ is composed by sampling the existing edges in the directed graph but the composition of the negative sample set $\mathcal{E}^{neg}$ is different from that in undirected graphs. Specifically, for each node $v$, $n$ nodes that do not point to $v$ are randomly sampled as source nodes to form negative samples with $v$. At the same time, $n$ nodes that are not pointed by $v$ are randomly sampled as target nodes to also form negative samples with $v$. We use the inner product and the sigmoid function as activation function to predict directed edges. The loss function is defined as the binary cross entropy loss:

\begin{equation}
	\mathcal{L} =   \mathbb{E}_{i \in \mathcal{E}^{pos}} - log \left( y^{pos}_{i}  \right) +  \mathbb{E}_{j \in \mathcal{E}^{neg}} - log \left( 1 - y^{neg}_{j}  \right),
	\label{eq9}
\end{equation}%
where $y^{pos}_{i}$ and $y^{neg}_{j}$ are the predicted value for the $i$-th positive samples and the $j$-th negative samples respectively. $W^{s}$ and $W^{t}$ in Eq.\ref{eq7} and Eq.\ref{eq8} are trained by minimizing the loss $\mathcal{L}$.

\section{Experiment}
In this section, we empirically evaluate and discuss the performance of different models including COBA, on six real-world datasets and on three tasks, i.e., directed link prediction, node classification and graph construction.
\subsection{Dataset}
Six publicly available datasets are used, which are graphs in different application scenarios.

\begin{table*}[htbp]
	\centering
	\resizebox{\textwidth}{!}
	{
	\begin{tabular}{@{}ccccccc@{}}
		\toprule
		Dataset        & \#Nodes & \#Edges & Avg. degree & 0-indegree & 0-outdegree & \#Labels \\ \midrule
		Jung           & 6,120   & 50,535  & 16.51       & 63.92$\%$  & 1.06$\%$    & -        \\
		Wikivote       & 7,115   & 103,689 & 29.15       & 66.54$\%$  & 14.13$\%$   & -        \\
		Google         & 15,763  & 171,206 & 21.72       & 0.01$\%$   & 21.02$\%$   & -        \\
		Cora           & 23,166  & 91,500  & 7.90        & 40.09$\%$  & 8.48$\%$    & 10       \\
		Amazon-Photo   & 7,650   & 143,663 & 37.60       & 2.59$\%$   & 10.99$\%$   & 8        \\ 
		Pubmed         & 19,717  & 44,338  & 4.50        & 10.38$\%$  & 80.34$\%$   & 3        \\ \bottomrule
	\end{tabular}
}
	\caption{Statistics of datasets.}
	\label{table1}
\end{table*}

\emph{Jung}$\footnote{http://konect.cc/networks/subelj\_jung-j/}$ is a software dataset of JUNG 2.0.1 libraries where nodes represent  Java classes and edges denote the dependence between Java classes. \emph{Wikivote}$\footnote{http://snap.stanford.edu/data/wiki-Vote.html}$ is a social network where nodes represents users and edges describes the voting relationship between users. \emph{Google}$\footnote{http://konect.cc/networks/cfinder-google/}$ collects web links from the Google website where nodes are websites and edges represent hyperlinks. \emph{Cora}$\footnote{http://konect.cc/networks/subelj\_cora/}$ and \emph{Pubmed}$\footnote{https://linqs.org/datasets/\#pubmed-diabetes}$ are both citation networks with node labels. Nodes are academic papers, edges represent citation relationships and labels are categories of papers. \emph{Amazon-Photo}$\footnote{https://pytorch-geometric.readthedocs.io/en/latest/\_modules \\ /torch\_geometric/datasets/amazon.html}$ is a co-purchase graph. Nodes are products, edges describe the dependence of goods being purchased at the same time and labels indicate the category of goods. Table~\ref{table1} summarizes the details of all datasets, each with a specific scale and sparsity.

\subsection{Comparison Methods and their Settings}
The comparison methods include: (1) baseline undirected graph embedding methods, including random walk-based methods like DeepWalk~\cite{perozzi2014deepwalk}, LINE-1~\cite{tang2015line}, node2vec~\cite{grover2016node2vec}, and two spatial-based graph convolution methods GraphSAGE~\cite{hamilton2017inductive} and GAT~\cite{velickovic2017graph}; (2) directed graph embedding methods, highlighting on seven dual-embedding methods, i.e. LINE-2~\cite{tang2015line}, HOPE~\cite{ou2016asymmetric}, APP~\cite{zhou2017scalable}, NERD~\cite{khosla2019node}, ATP~\cite{sun2019atp}, DGGAN~\cite{zhu2021adversarial} and DiGAE~\cite{kollias2022directed}.

Among the undirected graph embedding methods, DeepWalk, LINE-1 and node2vec treat the input graph as a directed graph and generate walk paths through directed edges; GraphSAGE and GAT update the embeddings of central nodes by aggregating embeddings of source neighbors and these embeddings can be later aggregated by their target neighbors. We use the inner product of node embeddings to predict edges. For DeepWalk, node2vec and APP, the number of walks, the walk length and the window size are set as 10, 80 and 10 respectively. LINE-1 and LINE-2 are LINEs preserving first-order proximity and second-order proximity respectively. In LINE-2, vertex embeddings are regarded as source embeddings and context embeddings are regarded as target embeddings. We use identity matrix instead of attribute matrix in DiGAE for fair comparison, even though one-hot encoding could conceptually hinder the identification of directionality~\cite{kollias2022directed}. For COBA, we set 40 epochs to ensure the convergence of model training. The number of layers and the number of sampling neighbors are set to 1 and 2, respectively, and their sensitivity is further explored.

In the link prediction and graph reconstruction tasks, the embedding dimension of all methods is set to 128. In the node classification task, the embedding dimension of undirected graph embedding methods is still 128, while the dimensions of two embeddings in DGE methods are both set to 64 to obtain the final 128-dimensional concatenated embedding. All experimental results are the average of 10 runs on a Linux server with RTX 3090.

\subsection{Experiments on Link Prediction}
Different from link prediction in undirected graphs, the link prediction in directed graphs not only predicts whether there are edges between nodes, but also predicts the direction of edges. We randomly select 30$\%$ of the edges as test set and the remaining 70$\%$ as training set. Three compositions of negative samples are set for each test set. In the first setting, the negative samples are randomly sampled from the unconnected node pairs. In the second setting, 50$\%$ of the negative samples are from random sampling, and another 50$\%$ of the negative samples are obtained by reversing the unidirectional edges in the positive samples. In the third setting, all negative samples come from the opposite edges of the unidirectional edges in positive samples. Among them, the first setting is the same as that for undirected graph link prediction.

\begin{table*}[htbp]
	\centering
	\resizebox{\textwidth}{!}
	{
	\begin{tabular}{ccccccccccccc}
		\toprule
		\multirow{2}{*}{method} & \multicolumn{3}{c}{Jung}                                     & \multicolumn{3}{c}{Wikivote}                                  & \multicolumn{3}{c}{Google}                                 & \multicolumn{3}{c}{Cora}                                   \\ \cline{2-13} 
		& 0\%                & 50\%               & 100\%              & 0\%                & 50\%               & 100\%              & 0\%                & 50\%               & 100\%              & 0\%                & 50\%               & 100\%              \\ 
		\midrule
		DeepWalk          & 80.88   & 65.76   & 50.38   & 82.68   & 67.56   & 52.53   & 79.08   & 69.95   & 60.89   & \underline{92.60}   & 72.54   & 52.48          \\
		LINE-1            & 40.07   & 45.07   & 50.13   & 64.11   & 57.85   & 51.54   & 78.81   & 69.39   & 60.10   & 64.90   & 58.09   & 51.21         \\
		node2vec          & 92.94   & 71.72   & 50.33   & 88.27   & 70.47   & 52.67   & 84.08   & 69.68   & 55.43   & 81.85   & 66.60   & 51.51         \\ 
		GraphSAGE         & 74.52   & 62.19   & 50.26   & 58.84   & 54.95   & 50.44   & 71.58   & 63.08   & 55.46   & 48.46   & 49.26   & 50.20         \\
		GAT               & 80.31   & 65.31   & 50.20   & 77.75   & 64.68   & 51.58   & 85.69   & 71.30   & 58.60   & 85.72   & 68.80   & 52.04         \\ 
		\midrule
		LINE-2            & 59.34   & 54.74   & 50.10   & 87.15   & 69.66   & 51.98   & 74.66   & 65.40   & 56.29   & 68.11   & 59.44   & 50.77    \\
		HOPE              & 97.02   & 96.72   & 96.42   & 92.68   & 90.11   & 87.51   & \underline{95.64}   & 91.90   & 88.08   & 91.61   & 85.93   & 80.22         \\
		APP               & 89.16   & 88.79   & 88.41   & 73.37   & 66.86   & 60.40   & 85.45   & 85.75   & 85.83   & 79.77   & 75.93   & 72.29     \\ 
		NERD         	  & 57.17   & 56.93   & 56.70   & 76.70   & 74.34   & 72.06   & 69.28   & 66.07   & 62.80   & 84.00   & 82.95   & 81.90      \\
		ATP               & 96.28   & 97.80   & \underline{99.38}   & 85.07   & 90.65   & \textbf{96.43}   & 82.05   & 86.13   & 90.29   & 84.54   & 89.04   & \textbf{93.63} \\ 
		DGGAN             & \underline{97.79}   & \underline{97.97}   & 98.26   & \underline{97.68}   & \underline{95.35}   & \underline{95.05}   & 92.66   & \underline{92.62}   & \underline{93.11}   & 89.96   & \underline{90.22}   & 90.67   \\  
		DiGAE             & 88.77   & 81.43   & 74.06   & 94.67   & 77.26   & 59.88   & 88.46   & 83.63   & 78.85   & 83.30   & 81.21   & 79.18  \\
		\midrule
		COBA(ours)       & \textbf{98.75} & \textbf{98.99}   & \textbf{99.45}   & \textbf{98.90}   & \textbf{96.50}   & 94.75   & \textbf{99.35}   & \textbf{96.27}   & \textbf{94.56}   & \textbf{97.87}   & \textbf{94.26}   & \underline{91.53}  \\ 
		\bottomrule
	\end{tabular}
}
	\caption{AUC score of link prediction. 0$\%$,50$\%$ and 100$\%$ respectively represent the proportion of negative samples obtained by reversing unidirectional edges in test set. The best results are bolded and the second best results are underlined.}
	\label{table2}
\end{table*}

We use AUC score to evaluate the performance of all methods (see Table~\ref{table2}). It demonstrates that COBA achieves the best AUC value in most cases, which indicates its effectiveness for representing directed edges. Most of the comparison methods, including COBA do not perform as well in the third setting as in the first setting. This is because the third task is more difficult as the model needs not only to determine whether there is an edge but also to determine the direction of the edge. ATP generally performs better in the 100$\%$ setting than other settings, probably because it adds the hierarchical information of nodes when constructing the matrix, which is very helpful in identifying the edge directions. The undirected graph embedding methods usually achieve good results in the first setting (0$\%$), but their performance gradually declines as the proportion of directional judgment on edges increases, indicating that one embedding is difficult to preserve the directionality of edges.

\begin{table*}[htbp]
	\centering
	\resizebox{\textwidth}{!}
	{
	\begin{tabular}{ccccccc}
		\toprule
		\multirow{2}{*}{method} & \multicolumn{2}{c}{Amazon-Photo}                                     & \multicolumn{2}{c}{Pubmed}                                  & \multicolumn{2}{c}{Cora}        \\ \cline{2-7} 
		& Micro-F1                & Macro-F1               & Micro-F1              & Macro-F1                & Micro-F1               & Macro-F1   \\ 
		\midrule
		DeepWalk          & \underline{90.15}   & \underline{88.96}   & 44.03   & 32.98   & \underline{69.68}   & 59.64         \\
		LINE-1            & 87.02   & 87.15   & 47.95   & 40.23   & 46.10   & 25.09         \\
		node2vec          & 89.02   & 88.70   & 39.01   & 28.43   & 60.58   & 41.97         \\ 
		GraphSAGE         & 39.88   & 5.70    & 44.83   & 32.49   & 58.63   & 52.46    \\
		GAT               & 39.66   & 5.69    & 58.50   & 43.42   & 64.84   & 54.43   \\
		\midrule
		LINE-2            & 85.97   & 85.87   & 47.31   & 37.66   & 40.65   & 15.79    \\
		HOPE              & 39.30   & 16.40   & 58.84   & 49.36   & 40.89   & 8.48        \\
		APP               & 25.40   & 10.86   & 65.19   & 61.74   & 69.14   & \underline{59.86}     \\ 
		NERD         	  & 88.58   & 87.73   & \textbf{75.69}   & \textbf{74.19}   & 66.33   & 53.18       \\
		ATP               & 52.42   & 44.36   & 53.08   & 46.66   & 40.24   & 11.28 \\ 
		DGGAN             & 34.51   & 11.81   & 43.07   & 31.98   & 40.24   & 6.49  \\  
		DiGAE             & 74.77   & 64.02   & 45.17   & 35.31   & 40.73   & 7.86    \\
		\midrule
		COBA(ours)        & \textbf{92.16} & \textbf{92.07}   & \underline{71.52}   & \underline{67.91}   & \textbf{70.75}   & \textbf{63.31}   \\ 
		\bottomrule
	\end{tabular}
}
	\caption{Micro-F1 and Macro-F1 scores of node classification. The best results are bolded and the second best results are underlined.}
	\label{table3}
\end{table*}

\subsection{Experiments on Node Classification}
We then apply all methods to the node classification task to compare their ability to learn the distribution of label-related features. We employ three datasets with node labels: Amazon-Photo, Pubmed and Cora (see Table~\ref{table1}). The ratio of training set to test set is still 7:3.

Table~\ref{table3} summarizes the experimental results on Micro-F1 and Macro-F1 indicators. It demonstrates that COBA achieves the best performance on Amazon-Photo and Cora; it does not perform as well as NERD on Pubmed, but still significantly outperforms other methods. It is noted that Pubmed is much sparser than the other two datasets, while NERD can learn the influence of neighboring nodes through alternate paths, so as to better retain the label information of nodes, especially in sparse networks. Although the bi-aggregation strategy in COBA can also alleviate the problem of insufficient learning caused by network sparsity, it relies more on information propagation from neighbors, which is less effective than the random walk strategy of NERD. Overall, COBA's performance is much more stable across different types of datasets. We also observe that the node classification performance of undirected graph embedding methods is similar to or even better than that of some DGE methods, which indicates that the edge direction has no significant influence on the information propagation of node labels and one embedding can also represent label-related features of nodes. Some DGE methods such as HOPE and DGGAN can achieve good results on link prediction but perform poorly and unstable on node classification. The main reason is probably that these methods intend to learn the structural features by designing specific proximity metrics and losses, without learning the label-related features of nodes, so it is difficult to generalize to the node classification task.

\begin{table*}[htbp]
	\centering
	\resizebox{\textwidth}{!}
	{
	\begin{tabular}{ccccccccccccc}
		\toprule
		\multirow{2}{*}{method} & \multicolumn{3}{c}{Jung}                                     & \multicolumn{3}{c}{Wikivote}                                  & \multicolumn{3}{c}{Google}                                 & \multicolumn{3}{c}{Cora}                                   \\ \cline{2-13} 
		& 0\%                & 50\%               & 100\%              & 0\%                & 50\%               & 100\%              & 0\%                & 50\%               & 100\%              & 0\%                & 50\%               & 100\%              \\ 
		\midrule
		COBA-Re-Co       & 98.18  & 98.70  & 99.46  & 98.70  & 96.36  & 94.02  & 99.33  & 96.00  & 93.96  & 97.46  & 93.85  & 91.80  \\
		\midrule
		COBA-Re          & 98.46  & 98.84  & \textbf{99.51}  & 98.78  & \textbf{96.66}  & \textbf{94.82}  & 99.19  & 96.17  & 94.50  & 97.44  & 94.16  & \textbf{92.73}  \\
		\midrule		
		COBA              & \textbf{98.75} & \textbf{98.99}   & 99.45   & \textbf{98.90}   & 96.50   & 94.75   & \textbf{99.35}   & \textbf{96.27}   & \textbf{94.56}   & \textbf{97.87}   & \textbf{94.26}   & 91.53  \\ 
		\bottomrule
	\end{tabular}
}
	\caption{Results of an ablation study. COBA-Re denotes the COBA removing the reverse aggregation, and COBA-Re-Co denotes the COBA removing both reverse and collaborative aggregations. The best results are marked in bold.}
	\label{table4}
\end{table*}

\subsection{Experiments on Graph Reconstruction}
\begin{figure}[h]
	\centering
	\includegraphics[width=1\columnwidth]{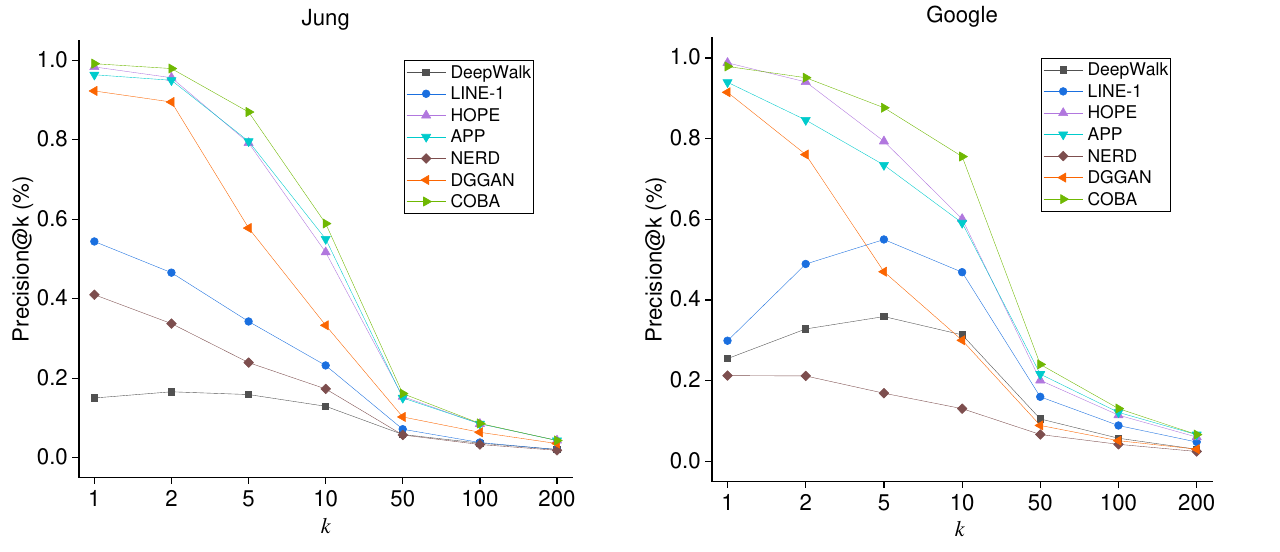}
	\caption{Precision@k of graph reconstruction.}
	\label{figure2}
\end{figure}

The graph reconstruction task is used to evaluate the quality of the embeddings in reconstructing the original graph. We conduct experiments on the Jung and Google datasets using the dot product of vectors to reconstruct the adjacent matrix. Without loss of generality, we randomly sample 10$\%$ of the nodes as the test set, where the top $k$ nearest target neighbors for each node are used to calculate the precision. We plot 7 representative methods for comparison in Figure~\ref{figure2}.

The experimental results show that COBA has the most powerful graph reconstruction ability under different $k$ values. It verifies that COBA can well preserve the neighborhood relationship through its sophisticated aggregation strategy, which is beneficial for graph reconstruction. As the value of $k$ increases, the precision of all methods gradually decreases, because the number of the neighbors of test nodes remain constant. Note that NERD achieves good results in both link prediction and node classification but does not perform well in graph reconstruction, even inferior to most undirected graph embedding methods. It suggests that the alternating walk strategy is not conducive to preserving the structural proximity of neighbors. For example, two nodes may not necessarily be structurally similar even though they point to the same node, but NERD considers them to be similar.

\subsection{Ablation Studies}
In this section, we investigate the impact of both bi-directional aggregation and collaborative aggregation on the performance of the COBA model. We observe the AUC changes in link prediction by removing different aggregation operations from the complete COBA. The ablation results on Jung, Wikivote, Google and Cora are shown in Table~\ref{table4}.

\begin{figure*}[t]
	\centering
	\includegraphics[width=1\columnwidth]{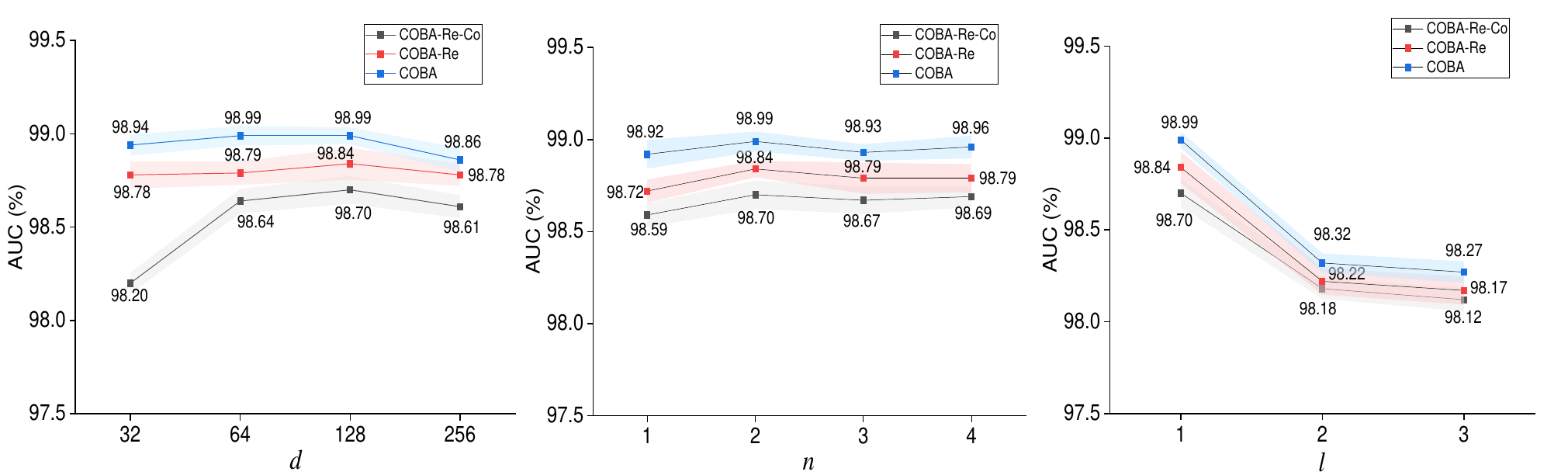}
	\caption{Results of parameter analysis.}
	\label{figure3}
\end{figure*}

The experimental results show that both bi-directional aggregation and collaborative aggregation are effective in most cases. However,  the bi-aggregation often fails as the proportion of directional judgment on edges increases. For example, COBA-Re slightly outperforms COBA on most datasets when the test set is composed of 100$\%$ reverse edges. It indicates that using information from neighbors in opposite direction is not beneficial to the direction judgment. Meanwhile, the model performance degrades slightly after using collaborative aggregation at the 0$\%$ setting on Google and Cora. The possible reason is that the two datasets are larger in scale, with more nodes and richer structures, while collaborative aggregation focuses more on local embedding correlations, making it difficult to learn the global structure of the graph well under the 0$\%$ setting. Note that COBA-Re-Co still outperforms existing methods, which shows that bi-directional aggregation can already learn the direction information from the source and target neighbors to update dual embeddings of central nodes.

\subsection{Parameter Analysis}
Finally, we analyze the influence of the parameters in COBA on the performance, including the embedding dimension $d$, the number of negative samples $n$ for each node and the number of aggregation layers $l$. We conducted link prediction task with the second setting (50$\%$) on the Jung dataset to observe the changes in model performance with different parameters, and the results are shown in Figure~\ref{figure3}.

Firstly, 32-dimensional embeddings are not sufficient to represent the properties of nodes, so the results of the model are the worst. With the increase of the dimension, the representative ability of node embeddings gradually improves and reaches the optimum at 128 dimensions. However, although the model can also achieve good results with 256 dimensions of embeddings, its performance is slightly lower than that with 128-dimensional embeddings, probably because the large-dimensional embeddings lead to over-fitting. 

Secondly, for different number of negative samples $n$, the model with $n=2$ achieves better result than that with $n=1$. As the value of $n$ increases, the performance tends to stabilize, but it brings additional computational burden. Therefore, $n$ is set to 2 by default as it is sufficient for training COBA. 

Finally, the model has the best performance when using one-layer aggregation. As the number of layers increases, the performance of the model decreases significantly, indicating that a deeper neural network will lead to the over-smoothing issue, making it difficult to distinguish the node representation.

\section{Conclusion}
In this paper, we propose a directed graph method based on collaborative bi-directional aggregation. This method can aggregate the information of neighbors from two directions, learn better representation for zero in/out-degree nodes, and correlate two embeddings by aggregating from each other. Extensive experiments on multiple datasets show that our method has promising performance in link prediction, node classification and graph reconstruction.






\bibliographystyle{elsarticle-num} 
\bibliography{mybibfile}

\begin{thebibliography}{10}
\expandafter\ifx\csname url\endcsname\relax
  \def\url#1{\texttt{#1}}\fi
\expandafter\ifx\csname urlprefix\endcsname\relax\def\urlprefix{URL }\fi
\expandafter\ifx\csname href\endcsname\relax
  \def\href#1#2{#2} \def\path#1{#1}\fi

\bibitem{grover2016node2vec}
A.~Grover, J.~Leskovec, node2vec: Scalable feature learning for networks, in:
  Proceedings of the ACM SIGKDD International Conference on Knowledge Discovery
  and Data Mining, 2016, pp. 855--864.

\bibitem{hamilton2017inductive}
W.~Hamilton, Z.~Ying, J.~Leskovec, Inductive representation learning on large
  graphs, in: Proceedings of the International Conference on Neural Information
  Processing Systems, Vol.~30, 2017, pp. 1024--1034.

\bibitem{perozzi2014deepwalk}
B.~Perozzi, R.~Al-Rfou, S.~Skiena, Deepwalk: Online learning of social
  representations, in: Proceedings of the ACM SIGKDD International Conference
  on Knowledge Discovery and Data Mining, 2014, pp. 701--710.

\bibitem{tang2015line}
J.~Tang, M.~Qu, M.~Wang, M.~Zhang, J.~Yan, Q.~Mei, Line: Large-scale
  information network embedding, in: Proceedings of the International
  Conference on World Wide Web, 2015, pp. 1067--1077.

\bibitem{khosla2019node}
M.~Khosla, J.~Leonhardt, W.~Nejdl, A.~Anand, Node representation learning for
  directed graphs, in: Joint European Conference on Machine Learning and
  Knowledge Discovery in Databases, Vol. 11906, 2019, pp. 395--411.

\bibitem{kollias2022directed}
G.~Kollias, V.~Kalantzis, T.~Id{\'e}, A.~Lozano, N.~Abe, Directed graph
  auto-encoders, in: Proceedings of the AAAI Conference on Artificial
  Intelligence, Vol.~36, 2022, pp. 7211--7219.

\bibitem{ou2016asymmetric}
M.~Ou, P.~Cui, J.~Pei, Z.~Zhang, W.~Zhu, Asymmetric transitivity preserving
  graph embedding, in: Proceedings of the ACM SIGKDD International Conference
  on Knowledge Discovery and Data Mining, 2016, pp. 1105--1114.

\bibitem{sun2019atp}
J.~Sun, B.~Bandyopadhyay, A.~Bashizade, J.~Liang, P.~Sadayappan,
  S.~Parthasarathy, Atp: Directed graph embedding with asymmetric transitivity
  preservation, in: Proceedings of the AAAI Conference on Artificial
  Intelligence, Vol.~33, 2019, pp. 265--272.

\bibitem{zhou2017scalable}
C.~Zhou, Y.~Liu, X.~Liu, Z.~Liu, J.~Gao, Scalable graph embedding for
  asymmetric proximity, in: Proceedings of the AAAI Conference on Artificial
  Intelligence, Vol.~31, 2017, pp. 2942--2948.

\bibitem{zhu2021adversarial}
S.~Zhu, J.~Li, H.~Peng, S.~Wang, L.~He, Adversarial directed graph embedding,
  in: Proceedings of the AAAI Conference on Artificial Intelligence, Vol.~35,
  2021, pp. 4741--4748.

\bibitem{cao2015grarep}
S.~Cao, W.~Lu, Q.~Xu, Grarep: Learning graph representations with global
  structural information, in: Proceedings of the ACM international Conference
  on Information and Knowledge Management, 2015, pp. 891--900.

\bibitem{wang2017community}
X.~Wang, P.~Cui, J.~Wang, J.~Pei, W.~Zhu, S.~Yang, Community preserving network
  embedding, in: Proceedings of the AAAI Conference on Artificial Intelligence,
  Vol.~31, 2017, pp. 203--209.

\bibitem{kipf2016semi}
T.~N. Kipf, M.~Welling, Semi-supervised classification with graph convolutional
  networks, in: International Conference on Learning Representations, 2017.

\bibitem{velickovic2017graph}
P.~Velickovic, G.~Cucurull, A.~Casanova, A.~Romero, P.~Lio, Y.~Bengio, Graph
  attention networks, in: International Conference on Learning Representations,
  2017.

\bibitem{wang2016structural}
D.~Wang, P.~Cui, W.~Zhu, Structural deep network embedding, in: Proceedings of
  the ACM SIGKDD International Conference on Knowledge Discovery and Data
  mining, 2016, pp. 1225--1234.

\bibitem{ma2019spectral}
Y.~Ma, J.~Hao, Y.~Yang, H.~Li, J.~Jin, G.~Chen, Spectral-based graph
  convolutional network for directed graphs, arXiv preprint arXiv:1907.08990
  (2019).

\bibitem{tong2020digraph}
Z.~Tong, Y.~Liang, C.~Sun, X.~Li, D.~Rosenblum, A.~Lim, Digraph inception
  convolutional networks, in: Proceedings of the International Conference on
  Neural Information Processing Systems, Vol.~33, 2020, pp. 17907--17918.

\bibitem{tong2020directed}
Z.~Tong, Y.~Liang, C.~Sun, D.~S. Rosenblum, A.~Lim, Directed graph
  convolutional network, arXiv preprint arXiv:2004.13970 (2020).

\bibitem{gopalakrishnan2020embedding}
S.~Gopalakrishnan, L.~Cohen, S.~Koenig, T.~Kumar, Embedding directed graphs in
  potential fields using fastmap-d, in: International Symposium on
  Combinatorial Search, Vol.~11, 2020, pp. 48--56.

\bibitem{madhavan2020directed}
R.~Madhavan, M.~Wadhwa, Directed graph representation through vector cross
  product, arXiv preprint arXiv:2010.10737 (2020).

\bibitem{salha2019gravity}
G.~Salha, S.~Limnios, R.~Hennequin, V.-A. Tran, M.~Vazirgiannis,
  Gravity-inspired graph autoencoders for directed link prediction, in:
  Proceedings of the ACM international Conference on Information and Knowledge
  Management, 2019, pp. 589--598.

\bibitem{bian2020rumor}
T.~Bian, X.~Xiao, T.~Xu, P.~Zhao, W.~Huang, Y.~Rong, J.~Huang, Rumor detection
  on social media with bi-directional graph convolutional networks, in:
  Proceedings of the AAAI Conference on Artificial Intelligence, Vol.~34, 2020,
  pp. 549--556.

\end{thebibliography}

\end{document}